\documentclass[english,aps,tightenlines,showpacs, showkeys, notitlepage]{revtex4-2}
\usepackage[T1]{fontenc}
\usepackage[latin9]{inputenc}
\usepackage{color}
\usepackage{babel}
\usepackage{amstext}
\usepackage{esint}
\usepackage[unicode=true,pdfusetitle,
 bookmarks=true,bookmarksnumbered=false,bookmarksopen=false,
 breaklinks=false,pdfborder={0 0 0},pdfborderstyle={},backref=false,colorlinks=true]
 {hyperref}
\begin{document}
\title{Torsion-Induced Modification to Friedmann Equations in $AdSL_{4}$
Gauged Gravity}
\author{Oktay Cebecio\u{g}lu$^{1}$}
\email{ocebecioglu@kocaeli.edu.tr}

\author{Salih Kibaro\u{g}lu$^{2}$}
\email{salihkibaroglu@maltepe.edu.tr}

\date{\today}
\begin{abstract}
We study the solution of the gravitational field equations in $AdSL_{4}$--gauged
gravity, a gauge-theoretic extension of general relativity based on
the $AdSL_{4}$ algebra. In this formulation, the antisymmetric gauge
field $B^{ab}$, associated with additional $AdSL_{4}$ tensorial
generators, induces space-time torsion via the relation $K^{ab}=\mu B^{ab}$,
where $K^{ab}$ denotes the contorsion 1-form. The presence of torsion
modifies both the spin connection and curvature, leading to an extended
set of Einstein--Cartan field equations. Focusing on spatially homogeneous
and isotropic cosmological backgrounds, we derive the modified Friedmann
equations which explicitly incorporate the torsional contribution.
The resulting acceleration equation admits de Sitter--like solutions
in which cosmic acceleration originates purely from the gauge--theoretic
structure of enlarged four-dimensional space-time symmetries. Within
this formulation, the dynamical components of the gauge field $B^{ab}$
emerge naturally as a source of the effective cosmological constants,
without the introduction of exotic matter sources. Furthermore, our
analysis shows that the torsion-driven cosmological phase in $AdSL_{4}$--gauged
gravity can reproduce an effective equation-of-state parameter $\omega_{B}=-1/3$,
establishing a connection between space-time torsion and cosmic-string--like
dynamics.
\end{abstract}
\affiliation{$^{1}$Department of Physics, Kocaeli University, 41380 Kocaeli, Turkey,}
\affiliation{$^{2}$Maltepe University, Faculty of Engineering and Natural Sciences,
34857, Istanbul, Turkey}
\maketitle

\section{Introduction}

Modern cosmology increasingly relies on gravitational theories that
extend general relativity (GR) to address phenomena such as early-universe
inflation, late-time cosmic acceleration, and possible deviations
from Einstein\textquoteright s theory at high energies. Such extended
theories of gravity, including higher-curvature models, scalar--tensor
theories, and gauge-algebra-based formulations, aim to capture geometric
or dynamical features of space-time that lie beyond the purely Riemannian
structure of GR \citep{Capozziello:2011Extended,Clifton:2011Modified,Nojiri:2017Modified,Bars:2013LocalConf,Dereli:2019GravitationalPlane,Ghilencea:2021GaugingScale,Dereli:2022DarkRange}. 

Among these approaches, gauge-theoretic formulations of gravity provide
a natural framework in which space-time geometry emerges from symmetry
principles \citep{Yang:1954Conservation,Utiyama:1956Invariant,Kibble:1961LorentzInvariance}.
A particularly interesting case is Maxwell--gauged gravity \citep{azcarraga2011generalized},
based on the Maxwell algebra \citep{bacry1970group,schrader1972maxwell,soroka2005tensor},
which augments the Poincaré symmetry with additional tensorial generators.
In their pioneering work \citep{azcarraga2011generalized}, the authors
developed a local four-dimensional gauge theory based on the Maxwell
algebra, applying it to generalize Einstein\textquoteright s gravity
through bilinear invariant curvature two-forms. While their action
is invariant under local Lorentz transformations, it does not preserve
full local Maxwell invariance; a fully Maxwell-symmetric formulation
was later proposed by \citep{Cardenas:2022GeneralizedEinstein} using
$AdSL_{4}$-valued gauge connections. By employing the Inönü--Wigner
contraction \citep{Penafiel:2018AdS,Cardenas:2022GeneralizedEinstein},
these constructions establish Maxwell gravity as a contraction limit
of $AdSL_{4}$-gravity. 

Previous studies have investigated the geometric and dynamical consequences
of this framework \citep{Gomis:2009Deformations,bonanos2010maxwell,durka2011gauged,soroka2012gauge,azcarraga2014minimalsuper,cebeciouglu2014gauge,cebeciouglu2015maxwell,concha2015generalized,Penafiel:2018AdS,kibarouglu2019maxwellSpecial,kibarouglu2019super,kibarouglu2020generalizedConformal,kibarouglu2021gaugeAdS,cebeciouglu2021maxwellMetricAffine,Cardenas:2022GeneralizedEinstein,Cebecioglu:2023MaxwellFR,Kibaroglu:2023GLSL}
as well as its cosmological solutions \citep{durka2011local,Azcarraga2013maxwellApplication,Kibaroglu:2022CosmoMaxwell,Kibaroglu:2024CosmMW}.
However, most of these works have either neglected or only partially
addressed the role of torsion, a manifestation of the antisymmetric
part of spacetime\textquoteright s affine connection, which can have
significant effects on gravitational dynamics, particularly in high-curvature
regimes relevant to the early universe.

In the development of cosmological models within the framework of
gauge theories of gravity, a central challenge lies in the consistent
incorporation of gauge fields. Various studies have explored gauge
field configurations in the Friedmann--Lemaître--Robertson--Walker
(FLRW) cosmology \citep{cervero1978classical,henneaux1982remarks,galt1991yang,moniz1993dynamics,bamba2008inflationary,gal2008non,maleknejad2013gauge,guarnizo2020dynamical}.
A particular line of research concerns non-Abelian Yang--Mills (YM)
theories, where the nonminimal coupling between YM fields and scalar
curvature has been investigated to assess its implications for cosmological
dynamics \citep{maleknejad2011non,sheikh2012gauge,maleknejad2013gauge,bielefeld2015cosmological}
(see also \citep{maleknejad2013gauge_report} for a comprehensive
review). Beyond its role in inflationary scenarios, YM theory has
been extensively applied in the context of dark energy \citep{Zhao:2005TheStateEquation,gal2008non,bamba2008inflationary,gal2012yang,elizalde2013cosmological,setare2013warm,rinaldi2015dark,mehrabi2017gaugessence}
and dark matter models \citep{zhang2010dark,buen2015non,gross2015non}.

In this work, we conduct an investigation of $AdSL_{4}$--gauged
gravity within the framework of FLRW metric. By explicitly incorporating
the torsional degrees of freedom induced by the additional tensorial
gauge contributions, we examine how these contributions modify the
effective gravitational field equations, influence the evolution of
the scale factor, and potentially offer novel mechanisms for realizing
cosmological solutions.

The paper is organized as follows. In Section (\ref{sec:Gauge-theory-of}),
we review the gauge theory construction based on the semi-simple extension
of the Poincaré group, commonly referred to as $AdSL_{4}$--gauged
gravity. We then introduce the gravitational action functional incorporating
the extended gauge fields. This is followed by a comprehensive geometric
analysis of torsion, highlighting its connection to the $AdSL_{4}$
gauge field and its influence on the spin connection and curvature.
Section (\ref{sec:Cosmological-setup}) develops the cosmological
setup by specializing the field equations to homogeneous and isotropic
space-times and explores corresponding Friedmann equations. Finally,
Section (\ref{sec:Conclusion}) summarizes our main results and discusses
potential directions for future work.

\section{Gauge theory of the $AdSL_{4}$ group\label{sec:Gauge-theory-of}}

We begin by providing an overview of the semi-simple extension of
the Poincaré group, originally introduced by \citep{soroka2012gauge}.
The algebra associated with this extension was independently re-derived
by \citep{Gomis:2009Deformations} through a deformation of the Maxwell
algebra. In contemporary literature, this structure is often referred
to as the $AdS$--Lorentz algebra \citep{durka2011gauged}. More
specifically, the $AdSL_{4}$ algebra can be systematically constructed
from the $AdS$ algebra via the $S$-expansion procedure, as developed
by \citep{Izaurieta:2006ExpandingLie} and further elaborated by \citep{Salgado:2014soD1}.
The fundamental commutation relations defining this algebra are given
by

\begin{eqnarray}
\left[M_{ab},M_{cd}\right] & = & i\left(\eta_{ad}M_{bc}+\eta_{bc}M_{ad}-\eta_{ac}M_{bd}-\eta_{bd}M_{ac}\right),\nonumber \\
\left[M_{ab},Z_{cd}\right] & = & i\left(\eta_{ad}Z_{bc}+\eta_{bc}Z_{ad}-\eta_{ac}Z_{bd}-\eta_{bd}Z_{ac}\right),\nonumber \\
\left[Z_{ab},Z_{cd}\right] & = & i\mu\left(\eta_{ad}M_{bc}+\eta_{bc}M_{ad}-\eta_{ac}M_{bd}-\eta_{bd}M_{ac}\right),\nonumber \\
\left[P_{a},P_{b}\right] & = & i\lambda Z_{ab},\nonumber \\
\left[M_{ab},P_{c}\right] & = & i\left(\eta_{bc}P_{a}-\eta_{ac}P_{b}\right),\nonumber \\
\left[Z_{ab},P_{c}\right] & = & i\mu\left(\eta_{bc}P_{a}-\eta_{ac}P_{b}\right).\label{eq: algebra}
\end{eqnarray}
In this framework, the generators are defined as $X_{A}=\left\{ P_{a},M_{ab},Z_{ab}\right\} $,
where $P_{a}$ correspond to translations, $M_{ab}$ to Lorentz transformations,
and $Z_{ab}$ to the $AdSL_{4}$ symmetry. For dimensional consistency,
the constant $\lambda$ is related to the cosmological constant $\Lambda$
which will be discussed later. The tangent space metric is taken to
be $\eta_{ab}=\text{diag}\left(+,-,-,-\right)$. In contrast to the
conventional Maxwell algebra, the generators $Z_{ab}$ in this semi-simple
extension are tensorial but non-Abelian and closed into an angular
momentum generator as given in (\ref{eq: algebra}). A notable property
of the semi-simple Poincaré algebra is that the Maxwell algebra can
be recovered from it via the Inönü--Wigner contraction procedure.

Following the approach of \citep{azcarraga2011generalized}, we formulate
the theory in terms of differential forms. To construct an action
based on the semi-simple Poincaré algebra, we begin with the gauge
connection expressed as the 1-form $A(x)=A^{A}X_{A}$,

\begin{equation}
A(x)=e^{a}P_{a}-\frac{1}{2}\omega^{ab}M_{ab}+\frac{1}{2}B^{ab}Z_{ab},\label{eq: gauge fields}
\end{equation}
where the 16 gauge fields $A^{A}(x)=\left\{ e^{a},\omega^{ab},B^{ab}\right\} $
correspond to the generators $AdSL_{4}$ symmetry transformations,
respectively. The associated curvature 2-form is defined by 

\begin{eqnarray}
F\left(x\right) & =dA+\frac{i}{2}\left[A,A\right]= & F^{a}P_{a}-\frac{1}{2}R^{ab}M_{ab}+\frac{1}{2}F{}^{ab}Z_{ab},
\end{eqnarray}
where $F^{a}$, $R^{ab}$ and $F^{ab}$ denote the generalized torsion,
the Lorentz curvature 2-form, and the additional $AdSL_{4}$ curvature
2-form. These are explicitly given by

\begin{eqnarray}
F^{a} & = & de^{a}+\omega_{\,\,c}^{a}\wedge e^{c}-\mu B_{\,\,c}^{a}\wedge e^{c}\nonumber \\
 & = & T^{a}-\mu B_{\,\,c}^{a}\wedge e^{c}\nonumber \\
 & = & \widetilde{D}e^{a},\label{eq:torsion Fa}
\end{eqnarray}

\begin{eqnarray}
R^{ab} & = & d\omega^{ab}+\omega_{\,\,c}^{a}\wedge\omega^{cb},
\end{eqnarray}

\begin{eqnarray}
F^{ab} & = & dB^{ab}+\omega_{\,\,c}^{[a}\wedge B^{c|b]}-\mu B_{\,\,c}^{a}\wedge B^{cb}-\lambda e^{a}\wedge e^{b}\nonumber \\
 & = & DB^{ab}-\mu B_{\,\,c}^{a}\wedge B^{cb}-\lambda e^{a}\wedge e^{b}\nonumber \\
 & = & \widetilde{D}B^{ab}-\lambda e^{a}\wedge e^{b},\label{eq:maxwell curv Fab}
\end{eqnarray}
where $T^{a}=De^{a}$ represents the ordinary torsion term, $D=d+\omega$
denotes the Lorentz exterior covariant derivative, while $\widetilde{D}=d+\widetilde{\omega}$
is defined with respect to the shifted connection 
\begin{eqnarray}
\widetilde{\omega}^{ab} & = & \omega^{ab}-\mu B^{ab}.\label{eq: shifted_connection}
\end{eqnarray}
The shifted connection thus provides a natural extension of the Riemannian
spin connection $\omega^{ab}$ to a non-Riemannian geometry with torsion.
In this sense, the resulting framework may be viewed as an Einstein--Cartan-type
\citep{Kibble:1961LorentzInvariance,Sciama1962analogy} geometry determined
by the algebraic structure of the $AdSL_{4}$ algebra. Importantly,
since the gauge field $B^{ab}$ is antisymmetric ($B^{ab}=-B^{ba}$),
the symmetric part of the connection is absent. Consequently, the
non-metricity tensor vanishes, and the resulting geometry is torsional
but metric-compatible \citep{azcarraga2011generalized}.

Within this framework, we introduce the shifted curvature as follows;

\begin{equation}
J^{ab}=R^{ab}(\omega)-\mu F^{ab}\equiv\widetilde{R}^{ab}(\widetilde{\omega})+\mu\lambda e^{a}\wedge e^{b},\label{eq:shifted cur-1}
\end{equation}
where $\widetilde{R}^{ab}(\widetilde{\omega})=\widetilde{D}\widetilde{\omega}^{ab}$
denotes the Lorentz curvature 2-form constructed from the shifted
connection (\ref{eq: shifted_connection}). This modification is not
merely technical; it plays a central role in constructing an action
that remains strictly invariant under the subalgebra $\{M,Z\}$. By
encoding the contributions of the additional algebraic structure into
the curvature sector, the shifted curvature allows for a systematic
derivation of the gravitational dynamics associated with the algebra
(\ref{eq: algebra}), while the remaining generators define the gauge
structure through on-shell diffeomorphisms.

\subsection{Gravitational action}

In close analogy with the Einstein--Hilbert action, we consider the
pure gravitational action constructed from the shifted curvature \citep{Cebecioglu:2023MaxwellFR}:

\begin{equation}
S_{EHJ}=\frac{1}{2\kappa}\intop\mathcal{L}_{EHJ}
\end{equation}
where, upon promoting $R^{ab}\rightarrow J^{ab}$, the Lagrangian
4-form becomes

\begin{equation}
\mathcal{L}_{EHJ}=\frac{1}{2}\varepsilon_{abcd}J^{ab}\wedge e^{c}\wedge e^{d}=J^{ab}\wedge{}^{*}e_{ab}=J{}^{*}1.
\end{equation}
To simplify the notation, we introduce the basis $p$-forms defined
by the wedge product of vielbeins as $e_{a_{1}...a_{p}}=e_{a_{1}}\wedge...\wedge e_{a_{p}}$.
Under this convention, the exterior products of basis 1-forms in 4D
are expressed as $e_{a}$, $e_{ab}=e_{a}\wedge e_{b}$ and $e_{abc}=e_{a}\wedge e_{b}\wedge e_{c}$,
etc. The Hodge star duals of these forms are defined as: $^{*}e_{a_{1}...a_{p}}=\frac{1}{(4-p)}\varepsilon_{a_{1}...a_{4}}e^{a_{(p+1)}...a_{4}}$
and where $^{*}1=\frac{1}{4!}\varepsilon_{abcd}e^{abcd}$ denotes
the invariant volume element. This Lagrangian naturally includes the
Einstein--Hilbert term and the $AdSL_{4}$ contribution, formulated
within the framework of the shifted connection (\ref{eq:shifted cur-1}),
together with a cosmological constant term. 

To derive the field equations, one may proceed in two ways. The first
is to vary the Lagrangian independently with respect to the vierbein
$e^{a}$ and the connection 1-forms $\tilde{\omega}^{ab}$ (Palatini\textquoteright s
method), subsequently solving for the extended torsion. The second
approach is to impose the vanishing of extended torsion from the outset
and consider constrained variations. For Einstein--Hilbert-type Lagrangians
(linear in curvature) in the absence of matter, both methods yield
the same field equations, namely the source-free Einstein equations
with a cosmological term \citep{Olmo:2011uz}. In this work, we adopt
the first-order Palatini formalism, treating the dynamical variables
as ($e^{a}$, $\tilde{\omega}^{ab}$) as independent.

The variation of the action reads,

\begin{eqnarray}
\delta\mathcal{L}_{EHJ} & = & \delta J^{ab}\wedge{}^{*}e_{ab}+J^{ab}\wedge\delta{}^{*}e_{ab}\nonumber \\
 & = & \left(\widetilde{D}\delta\widetilde{\omega}^{ab}+\mu\lambda\delta e^{[a}\wedge e^{|b]}\right)\wedge{}^{*}e_{ab}+J^{ab}\wedge\delta e^{c}\wedge{}^{*}e_{abc}\nonumber \\
 & = & \widetilde{D}\delta\widetilde{\omega}^{ab}\wedge{}^{*}e_{ab}+\delta e^{c}\wedge\left[J^{ab}\wedge{}^{*}e_{abc}+6\mu\lambda{}^{*}e_{c}\right],\label{eq:var ehj}
\end{eqnarray}
where resulting expression leads to two sets of field equations. One
can write the Eq.(\ref{eq:var ehj}) up to a total exterior derivative
term and set the variation to zero: 

\begin{equation}
\delta\mathcal{L}_{EHJ}=\delta\widetilde{\omega}^{ab}\wedge F^{c}\wedge{}^{*}e_{abc}+\delta e^{c}\wedge\left[J^{ab}\wedge{}^{*}e_{abc}+6\mu\lambda{}^{*}e_{c}\right]=0,
\end{equation}
the first arises from variations with respect to the shifted connection
and gives the torsion equation:

\begin{equation}
F^{c}=0\Rightarrow T^{c}=\mu B_{\thinspace d}^{c}\wedge e^{d}.\label{eq:tor-B}
\end{equation}
This relation highlights that torsion emerges naturally from the gauge
fields $B^{ab}$. While the original connection $\omega^{ab}$ may
allow for torsion, the shifted connection $\widetilde{\omega}^{ab}$
is torsion-free on-shell. Hence, the antisymmetric field $B^{ab}$
acts as a genuine source of torsion (\ref{eq:tor-B}), even in vacuum
states, which will be discussed in the next subsection.

The second set of equations follows from variations with respect to
the vierbein and yields a generalized Einstein equation:

\begin{equation}
J^{ab}\wedge{}^{*}e_{abc}+6\mu\lambda{}^{*}e_{c}=0,
\end{equation}
and it can be decomposed as

\begin{equation}
R^{ab}\wedge{}^{*}e_{abc}+6\mu\lambda{}^{*}e_{c}=\mu F^{ab}\wedge{}^{*}e_{abc}.\label{eq:gen-eins}
\end{equation}
Recasting this relation into tensorial form by using the relations
$e_{a}^{\,\,\mu}e_{b}^{\,\,\nu}R^{ab}=\frac{1}{2}R{}_{\rho\sigma}^{\,\,\,\,\mu\nu}dx^{\rho}\wedge dx^{\sigma}$
and $e_{a}^{\,\,\mu}e_{b}^{\,\,\nu}F^{ab}=\frac{1}{2}F_{\rho\sigma}^{\,\,\,\,\mu\nu}dx^{\rho}\wedge dx^{\sigma}$,
we find

\begin{equation}
R_{\,\,\alpha}^{\mu}-\frac{1}{2}R\delta_{\,\,\alpha}^{\mu}-\Lambda\delta_{\,\,\alpha}^{\mu}=T_{B}{}_{\,\,\alpha}^{\mu},\label{eq:gen-eins-tensor}
\end{equation}
where the cosmological constant is $\Lambda=6\mu\lambda$ and the
stress-energy tensors of the $AdSL_{4}$ gauge field $B_{\mu}^{ab}$
is defined by
\begin{equation}
T_{B}{}_{\,\,\alpha}^{\mu}=\mu e_{a}^{\,\,\nu}e_{b}^{\,\,\mu}\left(D_{[\nu}B_{\alpha]}^{\,\,ab}-\mu B_{[\nu\,\,c}^{\,\,a}B_{\alpha]}^{\,\,cb}\right)-\frac{\mu}{2}\delta_{\,\,\alpha}^{\mu}e_{a}^{\,\,\gamma}e_{b}^{\,\,\kappa}\left(D_{[\gamma}B_{\kappa]}^{\,\,ab}-\mu B_{[\gamma\,\,c}^{\,\,a}B_{\kappa]}^{\,\,cb}\right),\label{eq: T_B}
\end{equation}
where, in our notation, the square brackets around the indices imply
antisymmetrization. Additionally, we define the stress-energy tensor
related to the cosmological constant as $T_{\Lambda}{}_{\,\,\alpha}^{\mu}=\Lambda\delta_{\,\,\alpha}^{\mu}$,
therefore the total stress-energy tensor can be written as $T_{\,\,\alpha}^{\mu}=T_{B}{}_{\,\,\alpha}^{\mu}+T_{\Lambda}{}_{\,\,\alpha}^{\mu}$. 

Thus, the Einstein--Hilbert--like dynamics are extended by a natural
contribution from the $AdSL_{4}$ field $B_{\,\,\,\mu}^{ab}$, with
the cosmological constant term arising intrinsically from the algebraic
structure of the theory, rather than being introduced by hand. In
this formulation, the entire matter--energy content of the Universe
is encoded in the $AdSL_{4}$ gauge fields.

\subsection{Geometric analysis of torsion }

Torsion is a fundamental geometrical quantity in gauge formulations
of gravity, measuring the failure of parallel transport to preserve
the position of a vector in the tangent space. In recent decades,
the role of torsion in gravitational theory has been extensively studied,
primarily in the context of advancing gravity toward a gauge-theoretic
formulation and incorporating spin within a geometric framework. For
the cosmological point of view, torsion can influence the dynamics
of the early universe by introducing additional effective energy--momentum
contributions, thereby becoming significant near Planckian regimes.
Such effects can regularize singularities, alter inflationary dynamics,
and leave imprints on primordial perturbations \citep{Shie:2008TorsionCosmology,Poplawski:2010CosmologyTorsion,Poplawski:2011Nonsingular,Cai:2016ftTeleparallel,Yan:2020Interpreting}.

Let $e^{a}=e_{\mu}^{a}dx^{\mu}$ be the orthonormal coframe (vierbein)
and $\omega^{ab}=\omega_{\mu}^{ab}dx^{\mu}$ the full spin connection
(1-forms). The torsion 2-form is defined by

\begin{equation}
T^{a}=De^{a}=de^{a}+\omega_{\,\,b}^{a}\wedge e^{b}.
\end{equation}
We decompose the full spin connection into the torsionless (Levi--Civita
/ Lorentz) part $\mathring{\omega}_{\,\,b}^{a}$ and the contorsion
1-form $K_{\,\,b}^{a}$:

\begin{equation}
\omega_{\,\,b}^{a}=\mathring{\omega}_{\,\,b}^{a}+K_{\,\,b}^{a}\label{eq: full_spin_connection}
\end{equation}
where $\mathring{\omega}_{\,\,b}^{a}$ can be expressed in terms of
the Christoffel symbols $\Gamma_{\,\,\mu\nu}^{\lambda}=\frac{1}{2}g^{\lambda\rho}\left(\partial_{\mu}g_{\nu\rho}+\partial_{\nu}g_{\mu\rho}-\partial_{\rho}g_{\mu\nu}\right)$
as

\begin{equation}
\mathring{\omega}_{\mu}^{ab}=e_{\nu}^{a}\partial_{\mu}e^{b\,\nu}+e_{\nu}^{a}\Gamma_{\mu\sigma}^{\nu}e^{b\,\sigma}.
\end{equation}
By definition $\mathring{\omega}_{\,\,b}^{a}$ satisfies $de^{a}+\mathring{\omega}_{\,\,b}^{a}\wedge e^{b}=0$,
hence we obtain the standard algebraic relation
\begin{equation}
T^{a}=K_{\,\,b}^{a}\wedge e^{b}.\label{eq: torsion_2}
\end{equation}
From the previous calculation (\ref{eq:tor-B}), we know $F^{a}=0$
then we get
\begin{equation}
T^{a}=\mu B_{\,\,c}^{a}\wedge e^{c}.\label{eq: torsion_3}
\end{equation}
Comparing (\ref{eq: torsion_2}) and (\ref{eq: torsion_3}) shows
that the contorsion 1-form is directly given by the $AdSL_{4}$ gauge
field:
\begin{equation}
K_{\,\,b}^{a}=\mu B_{\,\,b}^{a}.
\end{equation}
Hence, the $AdSL_{4}$--gauge field $B^{ab}$ is not just an auxiliary
object: it generates torsion and therefore modifies both the spin
connection and curvature built from $\omega=\mathring{\omega}+K$.
Therefore, the full spin connection can be written in terms of the
$AdSL_{4}$ gauge field
\begin{equation}
\omega_{\,\,b}^{a}=\mathring{\omega}_{\,\,b}^{a}+\mu B_{\,\,b}^{a}.
\end{equation}
So the full $AdSL_{4}$ spin connection becomes

\begin{equation}
\omega_{\mu}^{ab}=e_{\nu}^{a}\partial_{\mu}e^{b\,\nu}+e_{\nu}^{a}\Gamma_{\mu\sigma}^{\nu}e^{b\,\sigma}+\mu B_{\mu}^{ab}.\label{eq: full_spin_connection_B}
\end{equation}
Furthermore, the Lorentz curvature 2-form of the full connection $\omega$
is
\begin{equation}
R^{ab}\left(\omega\right)=d\omega^{ab}+\omega_{\,\,c}^{a}\wedge\omega^{cb},
\end{equation}
and using (\ref{eq: full_spin_connection}) yields the decomposition
\begin{equation}
R^{ab}\left(\omega\right)=R^{ab}\left(\mathring{\omega}\right)+\mathring{D}K^{ab}+K_{\,\,c}^{a}\wedge K^{cb},
\end{equation}
where $\mathring{D}$ is the covariant derivative w.r.t. $\mathring{\omega}$.
Torsion thus affects the curvature dynamically and altering the Einstein
tensor, consequently, the gravitational field equations. Within this
framework, the gauge field $B^{ab}$ acts as the geometrical source
of torsion, which in turn provides a natural mechanism for these phenomena
within a gauge-theoretic formulation of gravity.

Finally, we remark that upon considering the shifted connection $\tilde{\omega}_{\,\,b}^{a}$
(\ref{eq: shifted_connection}), it becomes evident that this connection
coincides with the Levi-Civita connection, $\tilde{\omega}_{\,\,b}^{a}=\mathring{\omega}_{\,\,b}^{a}$.
This identification immediately implies a torsion-free configuration,
as ensured by the full torsion equation (\ref{eq:torsion Fa}). Consequently,
the extended torsion, constructed from the shifted connection, vanishes
identically, i.e., $\tilde{D}e^{a}=0$. In this sense, the shifted
framework effectively eliminates the extended torsional contributions,
thereby reducing the theory to a purely metric-compatible formulation.

\section{Generalized Friedmann Equations\label{sec:Cosmological-setup}}

The Einstein--Yang--Mills cosmology is a theoretical framework that
unifies general relativity with Yang--Mills theory to describe the
large-scale dynamics of the Universe. Within this approach, gravitation
is formulated as a connection between space-time and a non-Abelian
gauge field. Non-Abelian gauge fields have been extensively employed
to model the Universe during its early stages, particularly in scenarios
involving inflation and the subsequent formation of large-scale structures
such as galaxies and galaxy clusters. Motivated by these studies,
we investigate the possible cosmological implications of the gauge
fields $B_{\mu}^{ab}\left(x\right)$, which explicitly appear in the
gravitational field equations (\ref{eq:gen-eins-tensor}).

In the present cosmological context, we adopt the perfect fluid approximation,
which is motivated by the observed large-scale homogeneity and isotropy
of the Universe, consistent with the cosmological principle, and allows
the $AdSL_{4}$ curvature sector to be interpreted as an effective
matter--energy source. In this case, the total stress-energy tensor
is taken to be

\begin{equation}
T_{\,\,\alpha}^{\mu}=diag\left(\rho\left(t\right),-P\left(t\right),-P\left(t\right),-P\left(t\right)\right),
\end{equation}
where $\rho$$\left(t\right)$ and $P\left(t\right)$ denote the effective
energy density and isotropic pressure, respectively. These tensors
are consistent with the general form of a perfect fluid stress-energy
tensor, $T_{\mu\nu}=\left(\rho+P\right)u_{\mu}u_{\nu}-Pg_{\mu\nu}$
where $u_{\mu}$ is the fluid's four-velocity. Assuming again that
both $T_{B}{}_{\,\,\alpha}^{\mu}$ and $T_{\Lambda}{}_{\,\,\alpha}^{\mu}$
have perfect fluid form, the total energy density and pressure can
be written as

\begin{equation}
\rho=\rho_{B}+\rho_{\Lambda},\,\,\,\,\,\,\,\,P=P_{B}+P_{\Lambda},\label{eq: rho_P_total-1}
\end{equation}
where we consider all energy densities to be positive definite. 

Furthermore, each fluid component can be characterized by an effective
equation of state (EoS) parameter defined as $\omega_{B}=\frac{P_{B}}{\rho_{B}}$
and $\omega_{\Lambda}=\frac{P_{\Lambda}}{\rho_{\Lambda}}$, which
encapsulates the relation between pressure and energy density. In
the standard interpretation, $\omega_{\Lambda}=-1$ corresponds to
the cosmological constant\textquoteright s vacuum energy, while $\omega_{B}$
corresponds to the dynamics of the $B_{\mu}^{ab}\left(x\right)$ gauge
field. Moreover, the conservation of the stress-energy tensor in the
FLRW background is represented by the equation $\nabla_{\mu}T^{\mu\nu}=0$,
which consequently gives rise to the continuity equation,
\begin{equation}
\dot{\rho}+3H\left(\rho+P\right)=0,
\end{equation}
where $H\left(t\right)$ is the Hubble parameter.

In most cosmological models, the background geometry is assumed to
be spatially flat FLRW, consistent with observations indicating that
the Universe is nearly flat. Accordingly, we adopt the metric;

\begin{eqnarray}
ds^{2} & = & dt^{2}-a\left(t\right)^{2}\left(dx^{2}+dy^{2}+dz^{2}\right),\label{eq: FLRW metric}
\end{eqnarray}
where $a\left(t\right)$ is the scale factor. According to \citep{Azcarraga2013maxwellApplication,Kibaroglu:2022CosmoMaxwell,Kibaroglu:2024CosmMW},
the gauge fields $B_{\mu}^{ab}\left(x\right)$ can be expressed in
terms of two time-dependent fields $\psi\left(t\right)$ and $\zeta\left(t\right)$,

\begin{equation}
B_{\mu}^{0s}\left(x\right)\rightarrow B_{\mu}^{0s}\left(t\right)=\left(0,\delta_{i}^{s}\psi\left(t\right)\right),\label{eq: B_1}
\end{equation}

\begin{equation}
B_{\mu}^{rs}\left(x\right)\rightarrow B_{\mu}^{rs}\left(t\right)=\left(0,\epsilon_{i}^{\,\,rs}\zeta\left(t\right)\right),\label{eq: B_2}
\end{equation}
The ansatz for the gauge fields in Eqs. (\ref{eq: B_1}) and (\ref{eq: B_2})
is motivated by the requirement of spatial homogeneity and isotropy
in the FLRW sector. Specifically, for a $(1+3)$-dimensional space-time,
the gauge field components must be invariant under joint rotations
of the internal tangent indices and the $i,j,k$ spatial indices.
By identifying the $r,s$ tangent indices with the spatial indices
$i$ through the invariant $\mathrm{SO}(3)$ tensors $\delta_{i}^{s}$
and $\epsilon_{i}^{\,\,rs}$, with $\epsilon^{123}=+1$, the resulting
stress-energy tensor preserves the symmetries of the FLRW metric,
thereby allowing for a consistent cosmological reduction and the derivation
of the corresponding Friedmann equations. 

Substituting the ansatz Eqs.(\ref{eq: B_1}), (\ref{eq: B_2}), together
with the spin-connection expression (\ref{eq: full_spin_connection_B}),
into the gravitational field equations Eq.(\ref{eq:gen-eins-tensor}),
yields the corresponding Friedmann equations. From the $\left(0,0\right)$
component of Eq.(\ref{eq:gen-eins-tensor}), we obtain

\begin{equation}
\left(\frac{\dot{a}}{a}\right)^{2}=\frac{2\mu^{2}}{a^{2}}\left(\psi^{2}-\zeta^{2}\right)+\frac{\Lambda}{3},\label{eq: Friedmann_00}
\end{equation}
where a dot denotes differentiation with respect to cosmic time $t$.
The spatial diagonal $\left(i,i\right)$ components of Eq.(\ref{eq:gen-eins-tensor})
lead to
\begin{equation}
\frac{2\ddot{a}}{a}+\left(\frac{\dot{a}}{a}\right)^{2}=\frac{2\mu^{2}}{a^{2}}\left(\psi^{2}-\zeta^{2}\right)+\Lambda.\label{eq: Friedmann_ii}
\end{equation}
Combining these results, the acceleration equation takes the simple
form

\begin{equation}
\frac{\ddot{a}}{a}=\frac{\Lambda}{3}.\label{eq: Friedmann_acceleration}
\end{equation}
This result implies that the cosmic acceleration is constant and entirely
governed by the effective cosmological constant $\Lambda$ emerging
from the $AdSL_{4}$--gauged gravity framework. A positive $\Lambda$
leads to $\ddot{a}>0$, describing an exponentially expanding universe
$a\left(t\right)\propto e^{\sqrt{\Lambda}t}$, characteristic of both
de Sitter inflation like in the early universe and the current dark-energy--dominated
phase. A negative $\Lambda$, by contrast, would correspond to an
attractive effect that decelerates expansion and could even lead to
a recollapsing universe. So torsion-type contributions do not drive
acceleration; they only modify the expansion rate (the Hubble term).

Here the corresponding pressure and the energy density expressions
for the $B_{\,\,\,\mu}^{ab}$ gauge field are found as
\begin{equation}
P_{B}=-\frac{\mu^{2}}{4\pi Ga^{2}}\left(\psi^{2}-\zeta^{2}\right),\,\,\,\,\,\,\,\,\,\,\,\,\,\,\,\,\,\,\,\rho_{B}=\frac{3\mu^{2}}{4\pi Ga^{2}}\left(\psi^{2}-\zeta^{2}\right),\label{eq: pressure_energydensity}
\end{equation}
and considering the effective EoS for this gauge field, we get the
following expression,

\begin{equation}
\omega_{B}=\frac{P_{B}}{\rho_{B}}=-\frac{1}{3}.\label{eq: eos_B}
\end{equation}
The obtained value reveals that the $B_{\,\,\,\mu}^{ab}$ gauge field
behaves like a form of exotic matter with negative pressure. This
value lies between that of a pressureless dust ($\omega=0$) and a
cosmological constant ($\omega=-1$), but it is particularly noteworthy
because (\ref{eq: eos_B}) marks the critical threshold separating
decelerated and accelerated expansion in the Friedmann equations.
At this boundary, the contribution of the $B_{\,\,\,\mu}^{ab}$ field
neither strongly accelerates the Universe nor induces significant
deceleration, instead yielding a marginal effect on the expansion
rate. 

Furthermore, when we solve Eq.(\ref{eq: Friedmann_acceleration}),
the scale factor takes the following form;

\begin{equation}
a\left(t\right)=C_{1}e^{kt}+C_{2}e^{-kt},\label{eq: sol_a}
\end{equation}
where $k=\sqrt{\Lambda/3}$, $C_{1}$ and $C_{2}$ are the constants
of integration determined by the initial conditions of the system.
Substituting the solution (\ref{eq: sol_a}) into (\ref{eq: Friedmann_00}),
one derives the expression for the cosmological constant,
\begin{equation}
\Lambda=-\frac{3\mu^{2}\left(\psi^{2}-\zeta^{2}\right)}{2C_{1}C_{2}}.\label{eq: sol_Lambda}
\end{equation}
The result explicitly demonstrates that the scalar fields $\psi\left(t\right)$
and $\zeta\left(t\right)$ can be physically interpreted as the effective
source of the cosmological constant within this specific theoretical
framework. Additionally, from the definition of $\Lambda=6\lambda\mu$,
the relation between the parameters $\lambda$ and $\mu$ can be given
by
\begin{equation}
\lambda=-\frac{\mu\left(\psi^{2}-\zeta^{2}\right)}{4C_{1}C_{2}}.
\end{equation}

\section{Conclusion\label{sec:Conclusion}}

In this work, we examined the potential modification for the Friedmann
equations in the context of $AdSL_{4}$--gauged gravity. In this
formulation, an antisymmetric tensorial gauge field $B_{\,\,\,\mu}^{ab}$,
associated with the additional generators of the $AdSL_{4}$ algebra,
generates space-time torsion through the relation $T^{a}=\mu B_{\,\,c}^{a}\wedge e^{c}$
(\ref{eq: torsion_3}). Incorporating this torsion into the spin connection
(\ref{eq: full_spin_connection_B}) and curvature yields the $AdSL_{4}$
extended Einstein--Cartan equations. By specializing these equations
to spatially homogeneous and isotropic geometries, we derived their
cosmological sector, preserving all torsional contributions without
invoking additional matter sources or altering the underlying gauge-theoretic
structure.

By substituting the ansatz for the $AdSL_{4}$ gauge fields (\ref{eq: B_1})
and (\ref{eq: B_2}) together with the modified spin connection into
the generalized Einstein equations, we obtained a modified Friedmann
system in which the combination $\psi^{2}-\zeta^{2}$ of the gauge
field amplitudes appears in both the temporal ($00$) and spatial
($ii$) components. Eliminating this term between the two equations
yields an acceleration equation (\ref{eq: Friedmann_acceleration})
identical in form to that of a pure de Sitter universe, with acceleration
determined solely by the cosmological constant $\Lambda$. According
to the modified Friedmann equations given in (\ref{eq: Friedmann_00})
and (\ref{eq: Friedmann_ii}), the gauge field $B_{\,\,\,\mu}^{ab}$
influences the cosmic expansion history through its contribution to
the effective energy density in (\ref{eq: pressure_energydensity}).
Moreover, the scalar fields $\psi$ and $\zeta$, which constitute
the dynamical components of $B_{\,\,\,\mu}^{ab}$, arise as a source
of the cosmological constant $\Lambda$, thereby directly modifying
the cosmic acceleration rate, as shown in (\ref{eq: sol_Lambda}). 

The EoS parameter $\omega_{B}=-1/3$ (\ref{eq: eos_B}) obtained for
the $B_{\,\,\,\mu}^{ab}$ gauge field is particularly intriguing,
as it corresponds to a form of matter with a negative pressure that
scales as $P_{B}=-\frac{1}{3}\rho_{B}$. This specific value of the
EoS parameter is characteristic of a network of cosmic strings \citep{Kibble:1976Topology,Vilenkin:1981CosmicStrings,Vilenkin:1984CosmicStrings,Hindmarsh:1994CosmicStrings}
in the Universe, where the string tension effectively generates the
same dynamical behavior in the cosmological expansion. In the standard
cosmological context, such an EoS marks the critical boundary between
decelerating and coasting expansion, leading to a scale factor evolution
that is linear in cosmic time. The presence of a cosmic string--like
component, whether arising from fundamental field-theoretic defects
or emergent gauge-field dynamics as in our case, could have important
implications for structure formation, gravitational wave backgrounds
\citep{Damour:2000GravitationalWave,Caprini:2018Cosmological}, and
the thermal history of the early universe. This connection enriches
the physical interpretation of the $B_{\,\,\,\mu}^{ab}$ gauge field,
suggesting that its cosmological role might parallel that of topological
defects formed during symmetry-breaking phase transitions in the early
cosmos.

These results demonstrate that $AdSL_{4}$--gauged gravity with torsion
provides a mathematically consistent and physically motivated extension
of general relativity, capable of generating cosmic string-like components
in the cosmic energy budget from purely geometric origins. While the
late-time acceleration is entirely $\Lambda$-driven in the isotropic
background case, the torsional gauge fields may play a significant
role in intermediate epochs, spatial curvature dynamics, or anisotropic
settings \citep{WMAP:2003,WMAP:2006,WMAP:2008}. Future research directions
include constraining the model parameters $\mu$ and $\lambda$ using
cosmological observations, studying perturbations to assess effects
on structure formation and gravitational waves, and exploring whether
the curvature-like behavior of the $B_{\,\,\,\mu}^{ab}$ sector may
contribute to the resolution of cosmological singularities or provide
novel signatures of topological defects in a unified gauge--gravity
framework.

\section*{Declaration of competing interest }

The authors declare that they have no known competing financial interests
or personal relationships that could have appeared to influence the
work reported in this paper.

\section*{Declaration of generative AI and AI-assisted technologies in the
writing process}

During the preparation of this work the author used ChatGPT - GPT-4o
in order to improve readability of the text. After using this tool/service,
the author reviewed and edited the content as needed and takes full
responsibility for the content of the publication.

\section*{Data availability }

No data was used for the research described in the article.

\bibliography{maxwell_torsion_EH}

\end{document}